\documentclass[preprint,eqsecnum,aps,here,psfig]{revtex4}

\usepackage[dvips]{graphicx}

\newcommand{\ben}{\begin{eqnarray}}
\newcommand{\een}{\end{eqnarray}}
\makeatletter
\@addtoreset{equation}{section}
\makeatother
\begin{document}

\draft
\setlength{\baselineskip}{4mm}
\title{ NUCLEAR PHOTOABSORPTION AT PHOTON ENERGIES BETWEEN 300 AND 850 MEV }

\author{
Michihiro H{\small IRATA}\footnote{hirata@theo.phys.sci.hiroshima-u.ac.jp},
Nobuhiko K{\small ATAGIRI},
Kazuyuki O{\small CHI}}
\address{
Department of Physics, Hiroshima University, Higashi-Hiroshima 739, Japan}
\author{
and \\
Takashi T{\small AKAKI}\footnote{takaki@onomichi-u.ac.jp}}
\address{
Onomichi University, Onomichi 722-002, Japan}

\begin{abstract}
We costruct the formula for the
photonuclear total absorption cross section using the
projection method and the unitarity relation. Our treatment
is very effective when interference effects in the absorption
processes on a nucleon are strong.
The disappearance of the peak around the position of the
$D_{13}$ resonance in the nuclear photoabsorption can be explained
with the cooperative effect of the interference
in two-pion production processes,the Fermi motion,
the collision broadenings of $\Delta$ and $N^*$, and the pion distortion
in the nuclear medium. The change of the interference effect by the medium
plays an important role.
\end{abstract}
\maketitle
\setlength{\baselineskip}{24pt}
\newpage
\section{Introduction}
 The total photonuclearabsorption cross section has been measured over
broad mass number and in the energy range 300$\sim$1200 MeV at
Franscati\cite{Exp1,Exp2,Exp3}.Especially,it was noted that the
excitation peaks around the position of the$D_{13}$(1520 MeV)
and $F_{15}$(1680 MeV) resonances disappear and above 600 MeV
there is a strong reduction of the absolute value of
the cross section per nucleon compared with the data for hydrogen
\cite{Exp4,Exp5} and deutron\cite{Exp6}. These experimental findings
have been confirmed by the contemporary data on the photofission
cross section of $^{238}U$ \cite{Exp7}and$^{235}U$ \cite{Exp8} obtained
at Mainz up to 800 MeV. There were a couple of theoretical attempts to
explain the strong reduction of the cross section \cite{Phe1,Phe2,Phe3}.
It was necessary for them to assume large values for the
collision widths of the $D_{13}$ and $F_{15}$ resonances to explain
the above strong reduction \cite{Phe1,Phe2}. However,it was pointed
out in Ref.\cite{Phe3} that such significantly increasing resonance
widths were hardly justified. These theoretical analyses must miss
some important effects other than the collision broadening. In the
previous letter we pointed out that the disappearance of the peaks
around the position of the resonances higher than the $\Delta$ resonance
in the nuclear photoabsorption can be explained with
the cooperative effect of the interference in two-pion photoproduction
processes,the collision broadenings of $\Delta$ and $N^*$(1520),and the pion
distortion in the nuclear medium \cite{Theo4}. Our finding means
that the change of the interference effect by the medium plays
an important role.

 In this paper we present a detailed derivation of the formula for 
photonuclear total absorption cross sections based on the
projection operator technique and apply our method to evaluate them in the
extended energy region between 300 and 850 MeV. The experimental
total absorption cross sections on a nucleon show that
there is a fairly deep valley between the energy region of the $\Delta$
resonance and that of the $N^*$ resonance. On the other hand, the
experimental total cross sections on nuclei show that the above
valley is almost filled up. Furthermore, it is very interesting to note that
the mass number dependence of total cross sections quite vary with
the photon energy
as follows: $A^{0.8}$ around 300 MeV, $A^{1.7}$ around 500 MeV,
and $A^{0.65}$ around 750 MeV. We will discuss whether
the mass number dependence of total cross sections can be explained by our
theoretical treatment and speculate what one can learn from the
comparison of our results with the data.

 In section 2 we derive the formula for total nuclear photoabsorption
cross sections and in section 3, numerical results and discussions are
given. Section 4 contains our conclusions.

\section{ formalism of total nuclear photoabsorption cross section}
\label{sec:derivation}

Total photoabsorption cross section is propotional to the imaginary part
of
the elastic compton scattering T-matrix $T_{\gamma\gamma}$:
\ben
\sigma_{T}
&=& -\frac{2\Omega}{\upsilon}Im T_{\gamma\gamma
}\label{A1}\\
&=&i\frac{\Omega}{\upsilon}(T_{\gamma\gamma}-T_{\gamma\gamma}^{+}),
\label{A2}
\een
which is obtained from the unitarity relation.
Here $\Omega$ is the normalization
volume and $\upsilon$ is the relative velocity between the photon and
nucleus.
We will use this relation to
derive the formula of the total cross section. The expession for the
T-matrix
$T_{\gamma\gamma}$ is constructed using the projection operator technique
and focusing on the second resonance energy region where
the $N^*$(1520) resonance plays an important role
and the two-pion photoproduction prevails in addition to the one-pion
photoproduction.
The $N^*$(1520) resonance
can decay into both the $\pi$N and $\pi\pi$N channels, which branching
fractions are comparable, and its $\pi\pi$N decay occurs through two
dominant
modes, i.e., $\pi\Delta$ and $\rho$N. The two-pion
photoproduction takes place mainly through the $\pi\Delta$ and $\rho$N
channels.
So we include the
$N^*$,$\pi\Delta$ and $\rho$N intermediate states in our formalism
explicitly.
In order to simplify the formulation, we turn off the background
interactions except the $D_{13}$ channel in the one-pion production process,
which will be later added in the formalism.

We use the following projection operators to separate the nuclear
Hilbert
space into subspaces,
\begin{equation}
P_{\gamma}+p+q=1. \label{A3}
\end{equation}
$P_{\gamma}$ projects onto the space of photon plus nuclear ground
state, $p$
onto the spaces of both nuclear ground state and nuclear one particle-hole
states
except
the $P_{\gamma}$-space and $q$ onto the space of nuclear many particle-hole
states. We
assume that $P_{\gamma}$ and $q$ does not couple directly,
{\it i.e.\/},
$H_{\gamma q}=P_{\gamma}Hq=0$, and neglect higher order terms of
photo-coupling.  Here $H$ is the total Hamitonian of the system. Under
such
assumptions, the elastic compton scattering T-matix becomes
\begin{equation}
T_{\gamma\gamma}=H_{\gamma p}\frac{1}{E-{\cal{H}}_{pp}}H_{p\gamma},
\label{A4}
\end{equation}
with
\begin{equation}
{\cal{H}}=H+H\frac{q}{E-H_{qq}}H, \label{A5}
\end{equation}
where ${\cal{H}}_{pp}=p{\cal{H}}p$ and $H_{qq}=qHq$, and $E$ is the
total
energy of the system. The effective Hamiltonian $\cal{H}$ is
introduced so
as to eliminate the $q$-space.
The $p$-space is further divided into the following spaces:
\begin{equation}
p=P+D,\label{A6}
\end{equation}
with
\ben
P &=& P_{1}+P_{2},\label{A7}\\
D &=& D_{1}+D_{2},\label{A8}
\een
where $P_{1}$ projects onto the space of both one-pion plus nuclear
ground
state and one-pion plus nuclear one particle-hole states, $P_{2}$ onto the
space
of
two-pion plus nuclear particle-hole states, $\ D_{1}$ onto the space of one
$N^{\ast}$ plus nuclear one-hole states and $D_{2}$ onto the spaces of
both
$\pi\Delta$ plus nuclear one-hole states and $\rho$ plus nuclear particle-
hole states. Since $H_{\gamma q}=0$, then
${\cal{H}}_{\gamma P_{1}}=H_{\gamma P_{1}}$,
 ${\cal{H}}_{\gamma D_{1}}=H_{\gamma D_{1}}$ and
${\cal{H}}_{\gamma D_{2}}=H_{\gamma D_{2}}$.
Unlike the $\Delta$-hole model in the
pion-nucleus scattering, the $D_{1}$-space is not a doorway between the
subspaces
$P_{\gamma}$ and $P_{1}+D_{2}$, since the direct couplings described by
$H_{\gamma P_{1}}$ and $H_{\gamma D_{2}}$ are non-negligible.
$H_{\gamma P_{1}}$ corresponds to the background term in the $D_{13}$
channel and $H_{\gamma D_{2}}$ corresponds to the $\Delta$ and $\rho$
Kroll-Ruderman terms.
This fact
reflects the structure of the elastic compton scattering T-matrix which
will
be shown later. To simplify the evaluation of the T-matrix of
Eq.(\ref{A4}),
we must make some approximations for the reaction mechanism of the two-pion
production: (a) $P_{\gamma}$ and $P_{2}$ does not couple directly,
{\it i.e.\/}, $H_{\gamma P_{2}}=0$, so that ${\cal{H}}_{\gamma P_{2}}=0$.
(b) The transition between the space of $P_{1}+D_{1}$ and the space of
$P_{2}+D_{2}$ proceeds only from the direct coupling of $D_{1}$ and
$D_{2}$,so that
${\cal{H}}_{P_{1}P_{2}}={\cal{H}}_{P_{1}D_{2}}={\cal{H}}_{P_{2}D_{1}}=0$
, and ${\cal{H}}_{D_{1}D_{2}}=H_{D_{1}D_{2}}$. In
this
way, we assume that the $D_{2}$-space plays the role of the doorway
to two-pion states.

Inserting the projection operators of Eqs. (\ref{A6}), (\ref{A7}) and
(\ref{A8}) into Eq.(\ref{A4}) and using the above-mentioned
approximations, we
obtain the elastic compton scattering T-matrix given by
\begin{equation}
T_{\gamma\gamma}=T_{P_{1}}^{\gamma\gamma}+T_{D_{1}}^{\gamma\gamma}
                 +T_{D_{2}}^{\gamma\gamma}, \label{A9}
\end{equation}
with
\ben
T_{P_{1}}^{\gamma\gamma} &=& H_{\gamma P_{1}}G_{P_{1}}H_{P_{1}\gamma
},\label{A10}\\
T_{D_{1}}^{\gamma\gamma} &=& \tilde{F}_{\gamma
D_{1}}G_{D_{1}}F_{D_{1}\gamma}^{+},
\label{A11}\\
T_{D_{2}}^{\gamma\gamma} &=& H_{\gamma D_{2}}G_{D_{2}}H_{D_{2}\gamma},
\label{A12}
\een
The Green's functions in Eqs.(\ref{A10}), (\ref{A11}) and (\ref{A12})
are
defined as
\ben
G_{P_{1}} &=& (E-{\cal{H}}_{P_{1}P_{1}})^{-1},\label{A13}\\
G_{D_{2}}
&=& (E-H_{D_{2}D_{2}}-H_{D_{2}P_{2}}G_{P_{2}}^{0}H_{P_{2}D_{2}
}-\Sigma_{D_{2}})^{-1},\label{A14}\\
G_{D_{1}}
&=& (E-H_{D_{1}D_{1}}-H_{D_{1}P_{1}}G_{P_{1}}^{0}H_{P_{1}D_{1}
}-\Sigma_{D_{1}}-H_{D_{1}D_{2}}G_{D_{2}}H_{D_{2}D_{1}})^{-1},
\label{A15}
\een
where
\ben
\Sigma_{D_{1}} &=& H_{D_{1}q}\frac{1}{E-H_{qq}}H_{qD_{1}}+{\cal{H}}
_{D_{1}P_{1}}G_{P_{1}}{\cal{H}}_{P_{1}D_{1}}-H_{D_{1}P_{1}}G_{P_{1}}
^{0}H_{P_{1}D_{1}},\label{A16}\\
\Sigma_{D_{2}} &=& H_{D_{2}q}\frac{1}{E-H_{qq}}H_{qD_{2}}+{\cal{H}}
_{D_{2}P_{2}}G_{P_{2}}{\cal{H}}_{P_{2}D_{2}}-H_{D_{2}P_{2}}G_{P_{2}}
^{0}H_{P_{2}D_{2}}, \label{A17}
\een
and
\ben
G_{P_{1}}^{0} &=& (E-H_{P_{1}P_{1}})^{-1},\label{A18}\\
G_{P_{2}}^{0} &=& (E-H_{P_{2}P_{2}})^{-1}. \label{A19}
\een
The vertex functions in Eq.(\ref{A11}) are defined as
\ben
F_{D_{1}\gamma} &=& H_{D_{1}\gamma}+{\cal{H}}_{D_{1}P_{1}}G_{P_{1}}
H_{P_{1}\gamma}+H_{D_{1}D_{2}}G_{D_{2}}H_{D_{2}\gamma},\label{A20}\\
\tilde{F}_{\gamma D_{1}} &=& H_{\gamma D_{1}}+H_{\gamma P_{1}}G_{P_{1}
}{\cal{H}}_{P_{1}D_{1}}+H_{\gamma D_{2}}G_{D_{2}}H_{D_{2}D_{1}}.
\label{A21}
\een
In our formalism,  the background coupling of $P_{\gamma}$ and $P_{1}$
described by $H_{\gamma P_{1}}$ is included because the vertex correction
of the second term in eqs. (\ref{A20}) and (\ref{A21}) is
known to
be non-negligible in the elementary process. Since the strength of
$H_{\gamma
P_{1}}$ itself ,however,is small, we neglect the process of $P_{\gamma
}\rightarrow P_{1}\rightarrow q$. Thus Eq.(\ref{A10}) is approximately
written
as
\begin{equation}
T_{P_{1}}^{\gamma\gamma}\cong H_{\gamma
P_{1}}G_{P_{1}}^{0}H_{P_{1}\gamma},
\label{A22}
\end{equation}
and Eqs.(\ref{A20}) and (\ref{A21}) become
\ben
F_{D_{1}\gamma}^{+} &\cong & H_{D_{1}\gamma}+H_{D_{1}P_{1}}G_{P_{1}}^{0}
H_{P_{1}\gamma}+H_{D_{1}D_{2}}G_{D_{2}}H_{D_{2}\gamma},\label{A23}\\
\tilde{F}_{\gamma D_{1}} &\cong & H_{\gamma D_{1}}+H_{\gamma
P_{1}}G_{P_{1}
}^{0}H_{P_{1}D_{1}}+H_{\gamma D_{2}}G_{D_{2}}H_{D_{2}D_{1}}.
\label{A24}
\een
Hereafter, we use these approximate forms.

>From Eq.(\ref{A9}), we find the imaginary part of the T-matrix:
\begin{flushleft}
\ben
T_{\gamma\gamma}-T_{\gamma\gamma}^{+}
    &=& T_{P_{1}\gamma}^{+}\Delta G_{P_{1}}^{0}T_{P_{1}\gamma}+
T_{P_{2}\gamma}^{+}\Delta G_{P_{2}}^{0}T_{P_{2}\gamma}\nonumber\\
    & & { } +\Omega_{D_{1}\gamma}^{+}(\Sigma_{D_{1}}-\Sigma_{D_{1}}^{+})
\Omega_{D_{1}\gamma}
+\Omega_{D_{2}\gamma}^{+}(\Sigma_{D_{2}}-\Sigma_{D_{2}}^{+})
\Omega_{D_{2}\gamma},\label{A25}
\een
\end{flushleft}
with
\ben
\Omega_{D_{1}\gamma} &=& G_{D_{1}}F_{D_{1}\gamma}^{+}\label{A26}\\
\Omega_{D_{2}\gamma}
&=& G_{D_{2}}(H_{D_{2}\gamma}+H_{D_{2}D_{1}}G_{D_{1}
}F_{D_{1}\gamma}^{+}),\label{A27}
\een
and
\ben
T_{P_{1}\gamma} &=& H_{P_{1}\gamma}+H_{P_{1}D_{1}}\Omega_{D_{1}\gamma
},\label{A28}\\
T_{P_{2}\gamma} &=& H_{P_{2}D_{2}}\Omega_{D_{2}\gamma},\label{A29}
\een
where $\Delta G\equiv G-G^{+}.$ T-matrices of Eqs.(\ref{A28}) and
(\ref{A29})
describe the one-pion photoproduction and two-pion photoproduction,
respectively. The final state interaction, however, are not included in
these
expressions. Thus the cross sections calculated by Eqs.(\ref{A28}) and
(\ref{A29}) do not exactly correspond to experimental cross sections. In
our
formalism, the effect of the final state interaction is contained in the
third
and fourth terms of Eq.(\ref{A25}) and therefore there is an ambiguity
in
partitioning of the nuclear inelastic cross section. However, as far as
total
cross section is concerned, one can use the equation of (\ref{A25}) to
estimate it. \

In the Green's function $G_{D_{2}}$, the operator
$H_{D_{2}P_{2}}G_{P_{2}}%
^{0}H_{P_{2}D_{2}}$ is the free self-energies of the $\Delta$ and $\rho$
 meson
corrected by the Pauli-blocking effect. In order to clarify the physical

content of operators in the Green's function $G_{D_{1}}$, we rewrite it
as
\begin{equation}
G_{D_{1}}=(E-H_{D_{1}D_{1}}-H_{D_{1}P_{1}}G_{P_{1}}^{0}H_{P_{1}D_{1}}
-H_{D_{1}D_{2}}G_{D_{2}}^{0}H_{D_{2}D_{1}}-\Sigma_{D_{1}}-\Sigma_{D_{1}
}^{\prime})^{-1},\label{A30}
\end{equation}
where
\begin{equation}
\Sigma_{D_{1}}^{\prime}=H_{D_{1}D_{2}}(G_{D_{2}}-G_{D_{2}}^{0})H_{D_{2}D_{1}
},\label{A31}
\end{equation}
and
\begin{equation}
G_{D_{2}}^{0}=(E-H_{D_{2}D_{2}}-H_{D_{2}P_{2}}G_{P_{2}}^{0}H_{P_{2}D_{2}
})^{-1}.\label{A32}
\end{equation}
Here the operator $H_{D_{1}P_{1}}G_{P_{1}}^{0}H_{P_{1}D_{1}}$ consists
of both
the free $N^{\ast}$ self-energy due to the one-pion channel corrected by
the
Pauli-blocking effect and the pion rescattering term arising from the
coherent
$\pi^{0}$ production. The operator
$H_{D_{1}D_{2}}G_{D_{2}}^{0}H_{D_{2}D_{1}}$
corresponds to the free $N^{\ast}$ self-energy due to the two-pion
channel
corrected by the Pauli-blocking effect. The self-energies$
 \Sigma_{D_{1}}+$
$\Sigma_{D_{1}}^{\prime}$ and $\Sigma_{D_{2}}$ in Eqs.(\ref{A14}) and
(\ref{A30}) are complicated many-body operators arising from the
$q$-space
coupling as well as the $P_{1}$- and $P_{2}$-spaces coupling. The latter
coupling is related to the final state interaction in the one-pion or
two-pion
production. In practical calculations, these operators are assumed to be
simple one-body operators
\ben
\Sigma_{D_{1}}+\Sigma_{D_{1}}^{\prime} &\cong& W_{sp}^{(1)},\label{A33}\\
\Sigma_{D_{2}} &\cong& W_{sp}^{(2)},\label{A34}
\een
which are phenomenologically determined.

Using the expressions of Eq.(\ref{A25}) with Eqs.(\ref{A33}) and
(\ref{A34}),
the total cross section can be written as
\begin{flushleft}
\ben
\sigma_{T} &=& \frac{\Omega}{\upsilon}[\sum 2\pi\delta(E-H_{P_{1}P_{1}})
\left| T_{P_{1}\gamma}\right|^{2}+\sum 2\pi\delta(E-H_{P_{2}P_{2}})
\left| T_{P_{2}\gamma}\right|^{2}\nonumber\\
           & & { } +\Omega_{D_{1}\gamma}^{+}
(-2Im W_{sp}^{(1)})\Omega_{D_{1}\gamma}+\Omega_{D_{2}\gamma}^{+}
(-2Im W_{sp}^{(2)})\Omega_{D_{2}\gamma}\nonumber\\
           & &-\Delta_{1}-\Delta_{2}],
\label{A35}
\een
\end{flushleft}
where
\begin{flushleft}
\ben
\Delta_{1} &=& \sum 2\pi\delta(E-H_{P_{2}P_{2}})\left|
H_{P_{2}D_{2}}G_{D_{2}
}H_{D_{2}D_{1}}\Omega_{D_{1}\gamma}\right|  ^{2}\nonumber\\
           &-& \sum 2\pi\delta(E-H_{P_{2}P_{2}})
\left|  H_{P_{2}D_{2}}G_{D_{2}}^{0}
H_{D_{2}D_{1}}\Omega_{D_{1}\gamma}\right|  ^{2},\label{A36}\\
\Delta_{2} &=& \Omega_{D_{1}\gamma}^{+}H_{D_{1}D_{2}}G_{D_{2}}^{+}
(-2Im W_{sp}^{(2)})G_{D_{2}}H_{D_{2}D_{1}}\Omega_{D_{1}\gamma}.
\label{A37}\
\een
\end{flushleft}
Here the terms $\Delta_{1}$ and $\Delta_{2}$ appear due to the introduction
of
the one-body operator $W_{sp}^{(1)}$ and are subtracted in order to avoid
the
double counting of the processes included in $W_{sp}^{(1)}$.
Eq.(\ref{A35}) is our starting point
to calculate the total cross section. This equation shows that the
absorption
processes consist of three components, i.e.,the quasi-free processes such as
the one-pion photoproduction and two-pion photoproduction,
and genuine many-body absorption processes arising from the interaction
between the resonance (or pion) and the nucleon in a nucleus\cite{KMO}.
In actual calculation, the $\Delta$
excitation and the remaining background processes except the $D_{13}$
channel
in the one-pion photoproduction must be added in the above formula.

Now we turn to discuss how to evaluate the total cross section practically.
For the one-pion
photoproduction, the T-matrix is given in terms of the $\Delta$ and $N^*$
resonant amplitudes described by the isobar model and the remaining
background
multipole amplitudes. We employ
the model by Ochi et al.\cite{takaki} for the two-pion photoproduction.
For simplicity we use the Fermi gas model for a nucleus.

The cross section of
one-pion photoproduction off a proton in the nuclear matter is given in the
laboratory frame by
\begin{flushleft}
\begin{eqnarray}
 \sigma^{\pi}_{p}& =&
\frac{1}{v}\frac{3Z}{8\pi({k^p_f})^3}\int^{k^p_f}_0
d\vec{p}_1\int\frac{d\vec{q}}{(2\pi)^3}
\frac{d\vec{p}}{(2\pi)^3}(2\pi)^4
\delta^4(k+p_1-q-p)  \nonumber \\
&\;&\times\frac{1}{2}\sum_{\lambda,\nu,\nu^{\prime}}
\sum_{t_{\pi},t_{N}}
|<\vec{q}t_{\pi}\vec{p}\nu^{\prime}t_{N}|T_{P_{1}\gamma}|
\vec{k}\lambda\vec{p}_1\nu>|^2
\theta(|\vec{p}|-k^{t_{N}}_f)
\frac{M^2}{2k2\omega_{\vec{q}}E_{\vec{p}_1}
E_{\vec{p}}},\\
T_{P_{1}\gamma}&=&T_{N^{*}}+T_{\Delta}+T_{B},
\end{eqnarray}
\end{flushleft}
where
$T_{N^{*}}$ and $T_{\Delta}$ represent the $N^{*}$ and $\Delta$
resonance terms, respectively and $T_{B}$ is the background term.
${\vec{k}}$, ${\vec{p}_{1}}$, ${\vec{q}}$ and ${\vec{p}}$ are the
momenta of the incident photon, target proton, outgoing pion
and outgoing nucleon, respectively.
$E_{\vec{p}_{1}}$, $\omega_{\vec{q}}$, and $E_{\vec{p}}$ are the
energies of the target proton, outgoing pion, and outgoing nucleon,
respectively.
$Z$ and $v$ denote the proton number and the relative velocity between the
photon and nucleus, respectively.
$k_{f}^{t_{N}}$ is the Fermi momentum depending on the isospin quantum
number $t_{N}$.
The notation for all other spin-isospin quantum numbers are
self-explanatory.
The $N^{*}$ resonance term is expressed as
\begin{eqnarray}
T_{N^*}&=&F_{\pi NN^*}
G_{N^{*}}(s)
\tilde{F}^{+}_{\gamma PN^*},\\
G_{N^{*}}(s)&=&
               [\sqrt{s}-(M_{N^*}(s)+\delta M_{N^{*}})
                    +i(\Gamma_{N^*}(s)+\Gamma_{N^*sp})/2]^{-1},
\end{eqnarray}
where $\sqrt{s}$ is the total center of mass energy.
$M_{N^{*}}$ and $\Gamma _{N^{*}}$ in the $N^{*}$ propagator $G_{N^{*}}$
are the mass and the free width of $N^{*}$, respectively,
which are given so as to describe the energy dependence of the $\pi N$
$D_{13}$-wave scattering amplitude and the branching ratios at the
resonance energy. As the medium corrections, we introduce
the mass shift $\delta M_{N^{*}}$ and spreading width $\Gamma _{N^{*}sp}$
due to the collisions between $N^{*}$ and other nucleons.
$\tilde{F}_{\gamma P N^{*}}^{+}$ and $F_{\pi N N^{*}}$ are the
$\gamma P N^{*}$ and $\pi N N^{*}$ vertex functions, respectively,
of which detailed forms are given in Ref. \cite{ochi}, and the former
operator corresponds to $F_{D_{1}\gamma}^{+}$ in
Eq.(\ref{A23}) which includes the vertex correction.
The $\Delta$ resonance term $T_{\Delta}$ is written in a similar
form with $T_{N^{*}}$ but is important only at the low energy range
less than 500 MeV. The effective $\gamma P \Delta$ coupling constant
including the vertex correction
is obtained by the same way used in the fixing of the $\gamma P N^{*}$
coupling
constant.
Here the Born term with the cutoff form factor employed
in Ref.\cite{BL} is assumed as the background multipole amplitude.
The Pauli blocking effect for the $\Delta$ decay into $\pi N$
becomes non-negligible at low energies, since the probability of the
nucleon
being emitted with a small momentum increases compared with the energy
region
of the $N^*$ resonance\cite{KMO,hirata}.
\begin{figure}[t]
\begin{center}
\includegraphics[height=12cm,width=10cm]{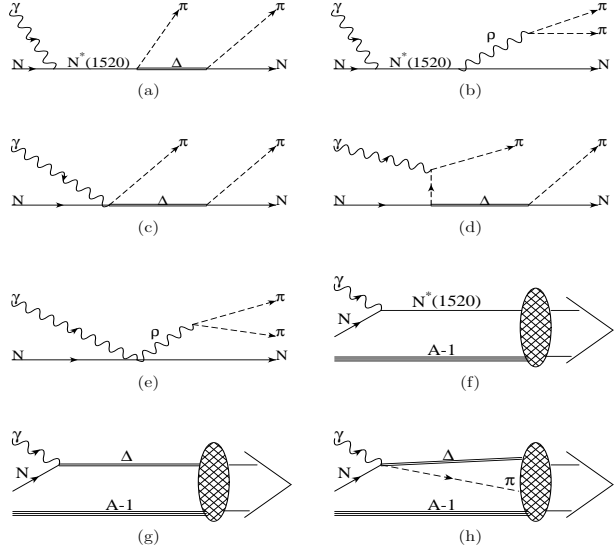}
\end{center}
\vspace{-4cm}
\caption{\small{Diagrams for the two-pion photoproduction on a nucleon
and genuine many-body absorption processes on a nucleus.
(a) The $N^{*} \rightarrow \pi \Delta$ contribution.
(b) The $N^{*} \rightarrow \rho N$ contribution.
(c) The $\Delta$ Kroll-Ruderman term.
(d) The $\Delta$ pion-pole term.
(e) The $\rho$ Kroll-Ruderman term.
(f) The many-body absorption process through the $N^{*}$.
(g) The many-body absorption process through the $\Delta$.
(h) The many-body absorption process through the $\pi \Delta$.
A is the mass number of the target nucleus.}}
\label{fig:fyman-diag}
\end{figure}
Thus, we include the Fock term in the $\Delta$ propagator
to modify the free $\Delta$ self-energy in addition to the collision
width.
The background amplitude $T_{B}$ is evaluated by using the experimental
multipole amplitudes
\cite{arndt}.
The integration over final particle momenta in Eq.(2.38) is performed by
using variables defined in the $\gamma N$ center of mass system.

The cross sections of two-pion photoproduction corresponding to
 Figs. \ref{fig:fyman-diag}(a),(b),(c),(d) and (e) are calculated
using the model of Ochi et al.\cite{takaki}.
The cross section of the two-pion photoproduction off a proton is given
by
\begin{eqnarray}
\sigma^{2\pi}_p&=&
                  \frac{1}{v}\frac{3Z}{8\pi({k^p_f})^{3}}
\int^{{k^p_f}}_0d\vec{p}_1
\int\frac{d\vec{q}_1}{(2\pi)^3}
\frac{d\vec{q}_2}{(2\pi)^3}
\frac{d\vec{p}}{(2\pi)^3}
(2\pi)^4\delta^4(k+p_1-q_1-q_2-p)
\nonumber
 \\
&\;\;\;&\times
\frac{1}{2}\sum_{\lambda,\nu,\nu^{\prime}}
\sum_{t_{{\pi}_1}t_{{\pi}_2}t_N}
|<\vec{q}_1t_{{\pi}_1}\vec{q}_2t_{{\pi}_2}\vec{p}t_N\nu^{\prime}|
T_{P_{2}\gamma}|\vec{k}\lambda\vec{p}_1\nu>|^2
\theta(|\vec{p}|-k^{t_{N}}_f)\nonumber\\
&\;\;\;&\times\frac{M^2}
{2k2\omega_{\vec{q}_1}2\omega_{\vec{q}_2}E_{\vec{p}_1}E_{\vec{p}}},
\end{eqnarray}
where ${\vec{q}_1}$ and ${\vec{q}_2}$ are the momenta of the outgoing
pions.
The medium-modified $T_{P_{2}\gamma}$ matrix is expressed as
\begin{eqnarray}
T_{P_{2}\gamma}&=&T_{\Delta KR}+ T_{\Delta PP}+ T^s_{N^*\pi\Delta}
   + T^d_{N^*\pi\Delta}+ T_{\rho {\rm KR}}+ T_{N^*\rho N}.
\end{eqnarray}
The $\Delta$ Kroll-Ruderman term is written as
\begin{eqnarray}
T_{\Delta KR}&=&
F_{\pi N\Delta}
G_{\pi \Delta}(s,\vec{p}_{\Delta})
F^{+}_{\Delta KR},
\\
G_{\pi \Delta}(s,\vec{p}_{\Delta})&=&
        [\sqrt{s}-\omega_{\pi}(\vec{p}_{\Delta})
-(M_{\Delta}(s,\vec{p}_{\Delta})+\delta M_{\Delta})
+i(\Gamma_{\Delta}(s,\vec{p}_{\Delta})+\Gamma_{\Delta sp})/2\nonumber\\
             & & { } -V_{\pi}(\vec{q}_{\pi})]^{-1} ,
\end{eqnarray}
Here $F_{\pi N\Delta}$
is the $\pi N \Delta$ vertex function,
and $F^{+}_{\Delta {\rm KR}}$
is the $\Delta$ Kroll-Ruderman vertex function.
The detailed forms of vertex functions are given in Ref. \cite{ochi}.
$V_{\pi}(\vec{q}_{\pi})$ is the pion self-energy due to the
distortion.
$\vec{p}_{\Delta}$ is the $\gamma N$ center of mass momentum of $\Delta$
and
$\vec{q}_{\pi}$ is the outgoing pion momentum.
$M_{\Delta}$ and $\Gamma_{\Delta}$ in the propagator $G_{\pi \Delta}$
are the mass and the free width of $\Delta$, respectively
and $\delta M_{\Delta}$ and $\Gamma_{\Delta sp}$ are
the mass shift and collision width, respectively.
Here we assume that the one-body operator $W_{sp}^{(2)}$ is
given by the sum of the pion optical potential and the $\Delta$ spreading
potential.
We neglect the medium correction for
the $\rho$ meson since it is far off-shell.
The other terms in the r.h.s. of Eq.(2.43) and the correction
term $\Delta_{1}$ are expressed in a similar way.
The detailed forms of free $T$ matrices are given in Ref. \cite{ochi}.

In addition to the one- and two-pion photoproduction
processes,
there are three genuine many-body processes which are shown
in Figs. \ref{fig:fyman-diag}(f),(g) and (h).
The cross section for Fig. \ref{fig:fyman-diag}(f) corresponds to
the third term of Eq.(\ref{A35}) and is given by
\begin{eqnarray}
\sigma^p_{N^*(A-1)}&=&\frac{1}{v}\frac{3Z}{8\pi (k^p_{f})^{3}}
\int^{k^p_{f}}_{0}d\vec{p}_1\Gamma_{N^*sp}
\frac{1}{2}\sum_{\lambda\nu\nu_{N^*}}|G_{N^*}(s)
<\vec{p}_{N^*}\nu_{N^*}|\tilde{F}^{+}_{\gamma PN^*}|\vec{k}\lambda
\vec{p}_1\nu>|^2\nonumber\\
&\;\;\;&\times\frac{1}{2k}\frac{M}{E_{\vec{p}_1}} .
\end{eqnarray}
The cross section for Fig. \ref{fig:fyman-diag}(g) has a similar form
with Eq.(2.46).
The cross section for Fig. \ref{fig:fyman-diag}(h) corresponds to
the fourth term of Eq.(\ref{A35}) and is given by
\begin{eqnarray}
\sigma^p_{\pi \Delta (A-1)}&=&\frac{1}{v}
                              \frac{3Z}{8 \pi{(k^p_{f})}^3}
\int^{k^p_{f}}_{0}
d\vec{p}_1\int\frac{d\vec{q}}{(2\pi)^3}
\frac{d\vec{p}}{(2\pi)^3}
(\Gamma_{\Delta sp}+2{\rm Im} V_{\pi}(\vec{q}))
(2\pi)^{3}\delta(\vec{k}+\vec{p}_{1}-\vec{q}-\vec{p})
\nonumber
\\
&\;\;&\times
\frac{1}{2}\sum_{\lambda\nu\nu_{\Delta}}|G_{\pi \Delta}(s,p_{\Delta})
<\vec{q}t_{\pi}\vec{p}\nu_{\Delta}t_{\Delta}|F^{+}
_{\gamma P\pi \Delta}|\vec{k}\lambda \vec{p}_1\nu>|^2
\frac{1}{2k2\omega_{\vec{q}}}\frac{M}{E_{\vec{p}_1}} ,
\end{eqnarray}
where
$F_{\gamma P \pi \Delta}^{+}$ describes the
$\gamma P \rightarrow \pi \Delta$ transition corresponding to
Figs. \ref{fig:fyman-diag} (a), (c) and (d).
To evaluate the cross section of Eq.(2.47), one needs to know
the momentum dependence of $\Gamma_{\Delta sp}$ and ${\rm Im}
V_{\pi}(\vec{q})$.
The width $\Gamma_{\Delta sp}$ is assumed to be constant in the
kinematical region where the process $\Delta$N $\rightarrow$ NN occurs and
zero outside this physical region.
The second term related to ${\rm Im} V_{\pi}(\vec{q})$ describes the process
that the $\Delta$ resonance decays into $\pi$N while the pion is absorbed by
nucleus.
In order to include this instability of the $\Delta$, we replace
${\rm Im}V_{\pi}(q)$
with ${\rm Im}\tilde{V}_{\pi}(\vec{q},E_{\Delta})$ which is
written as
\begin{eqnarray}
{\rm Im}\tilde{V}_{\pi}(\vec{q},E_{\Delta})=(-\frac{1}{\pi}
{\rm Im}\frac{1}{D(\sqrt{s_{\Delta}})}){\rm Im}V_{\pi}
(\vec{q},q_{0}),
\end{eqnarray}
where $s_{\Delta}=E_{\Delta}^{2}-\vec{p}^{2}$ and $q_{0}=k+E_{p_{1}}
-E_{\Delta}$, respectively and $D(\sqrt{s_{\Delta}})$ is the free D-function
of the $\Delta$. Then the expression obtained is integrated over the
$\Delta$ energy
$E_{\Delta}$ in addition to
momenta of  $\vec{p}_{1}$,$\vec{q}$ and $\vec{p}$. Generally, the pion
absorbed becomes off-shell, but in our
calculation we use one-shell value at $s_{\pi N}=q_{0}^{2}-\vec{q}^{2}$ for
${\rm Im}$ $V_{\pi}(\vec{q},q_{0})$ as an approximation.
The correction term $\Delta_{2}$ in Eq.(2.37) is evaluated in the
same way as Eq.(2.47).
The cross sections of photoabsorption on a neutron in the nuclear matter
are also given in a similar form with those of a proton.

\section{Numerical results and discussions}
\label{sec:Numerical}

 Let us start from the comments for photoabsorption reaction off a nucleon
because the information of the elementary pion productions is very
important to understand the strong damping mechanism of the $N^*$ resonance.
\begin{figure}[t]
\begin{center}
\includegraphics[height=12cm,width=12cm]{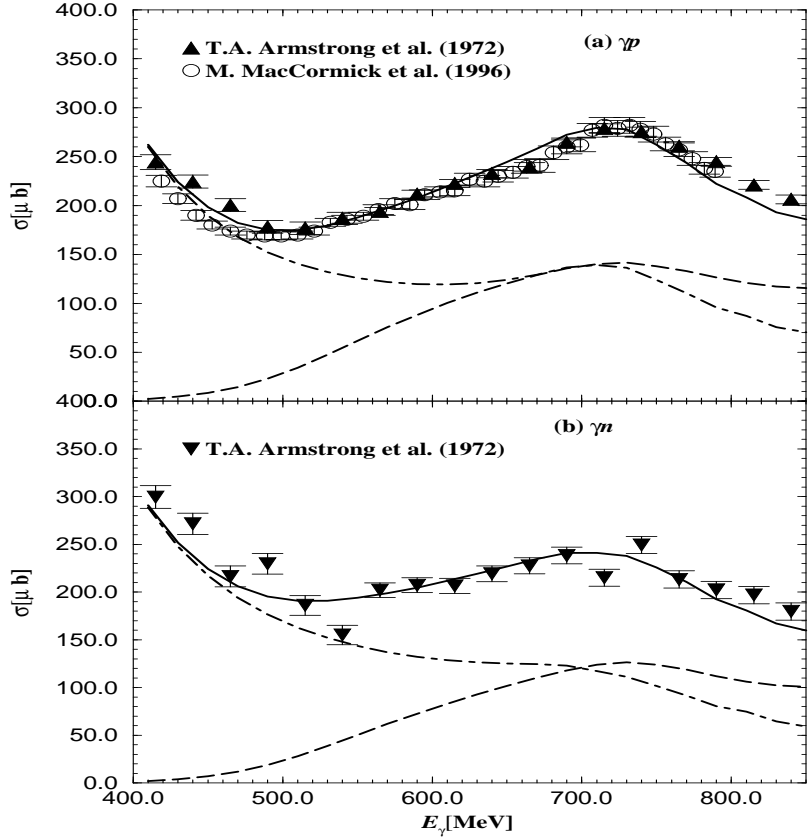}
\end{center}
\vspace{-1.5cm}
\caption{\small{
The total photoabsorption cross sections on a proton and a neutron.
The dash-dotted line is the contribution of the one-pion production
obtained by using SM95 amplitudes of
Arndt {\it et al.}\protect\cite{arndt}.
The dashed line is the contribution of the two-pion production
calculated by our model \protect\cite{ochi,takaki}. The solid line
is the sum of those contributions.
(a) Open circles and triangles (up) represent the data of total
photoabsorption
cross section on a proton\protect\cite{Exp5,Exp6}.
(b) Triangles (down) represent the data of
total photoabsorption cross section on a neutron\protect\cite{Exp6}.
}}
\label{fig:p-ntot}
\end{figure}
The dominant reactions on a nucleon over the photon energies from 300 to
850 MeV are the one-pion and two-pion photoproduction \cite{double-exp1}.
At first it is noted that the peak of the $N^*$ resonance energy region
shows up clearly by the combined effect of one-pion and two-pion productions
as shown in Fig. \ref{fig:p-ntot}(a) and (b), where
the cross section of one-pion
production has a small peak around 720 MeV and the cross section of
\begin{figure}[t]
\begin{center}
\includegraphics[height=14cm,width=8cm]{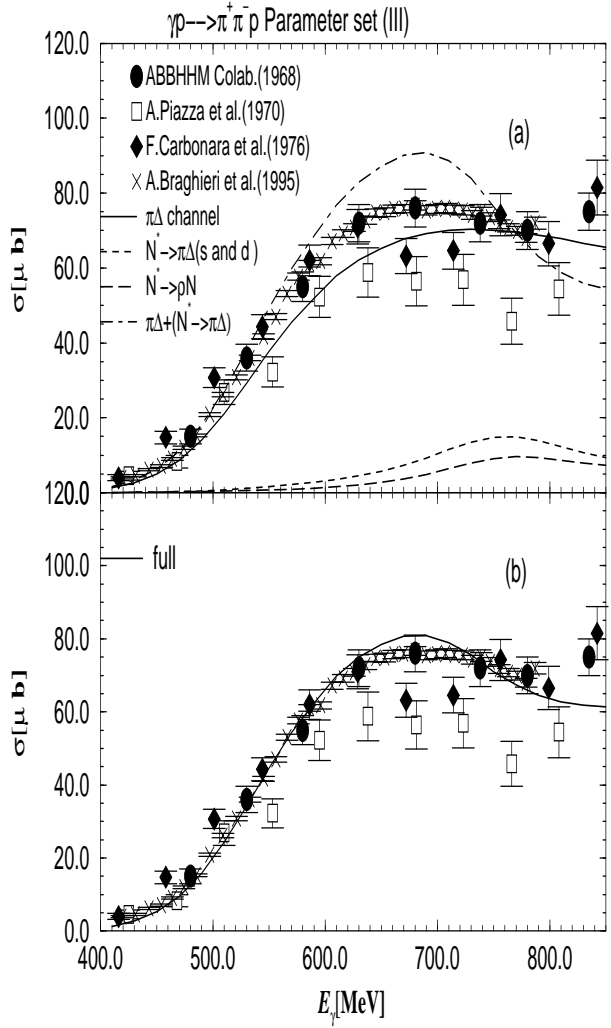}
\end{center}
\vspace{-1.0cm}
\caption{\small{
The total cross section for the $\gamma p \rightarrow
\pi^+\pi^-p$. (a) The solid line is the contributions of the
$\Delta$ Kroll-Ruderman and $\Delta$ pion-pole terms ($\pi \Delta$
channel), dashed line is the
contributions of the $N^* \rightarrow \pi \Delta$(s and d waves) terms,
long dashed line is the contribution of the $N^* \rightarrow \rho N $ term,
and dash-dotted line is the sum of contributions from $\pi \Delta$ channel
and
$N^* \rightarrow \pi \Delta$ term. (b) The solid line
corresponds to the full calculation. Theoretical lines are obtained
by using parameter set (III) in our model \protect\cite{takaki}.
Experimental data are taken
from Refs. \cite{double-exp1,Exp9,Exp10,Exp11}.
}}
\label{fig:pmp}
\end{figure}
two-pion production starts to grow from 400 MeV and increases up to around
800 MeV. The cross section of one-pion production is calculated by using
the amplitudes of Arndt et al. \cite{arndt} and that of two-pion
production is calculated by our model \cite{ochi,takaki}.

We briefly review
our model for the two-pion production.
\begin{figure}[t]
\begin{center}
\includegraphics[height=14cm,width=8cm]{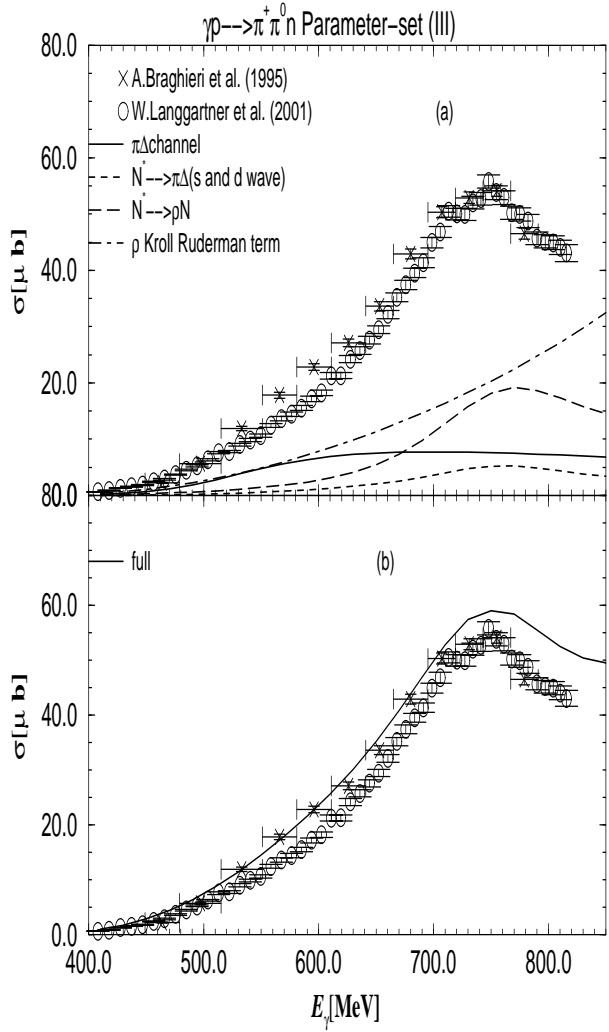}
\end{center}
\vspace{-1.0cm}
\caption{\small{
The total cross section for the $\gamma p \rightarrow
\pi^+\pi^0n$. (a) The solid line is the contributions of the
$\Delta$ Kroll-Ruderman and $\Delta$ pion-pole terms ($\pi \Delta$
channel), dashed line is the
contributions of the $N^* \rightarrow \pi \Delta$(s and d waves) terms,
long dashed line is the $N^* \rightarrow \rho N $ term, and dash-dotted
line is the contribution of $\rho$ Kroll-Ruderman term.
(b) The solid line
corresponds to the full calculation.
Theoretical lines are obtained
by using parameter set (III) in our model \protect\cite{takaki}.
 Experimental data are taken
from Refs. \protect\cite{double-exp1,double-exp3}.
}}
\label{fig:pzn}
\end{figure}
\vspace{1cm}
For the  $\gamma N \rightarrow \pi^{+} \pi^{-} N$ reaction,
four processes expressed by the diagrams
(a), (b), (c), and (d) in Fig. \ref{fig:fyman-diag} are assummed to
contribute to this channel. In these processes,
the $\Delta$ Kroll-Ruderman ($\Delta$KR) term [Fig.1(c)]
and the $\Delta$ pion-pole ($\Delta$PP) term [Fig.1(d)] dominate
on the cross section.
The $N^*$ contributions
alone are small, but the interference among the $N^*$ terms, the
$\Delta$KR and $\Delta$PP terms is important as shown in
Figs. \ref{fig:pmp}(a) and (b).
 Because of this, the $N^*$ excitation is regarded as an important
ingredient in the two-pion photoproduction.
For the $\gamma p \rightarrow \pi^{+} \pi^{0} n $ and
$\gamma n \rightarrow \pi^{-} \pi^{0} p$
reactions, the $\rho$ meson Kroll-Ruderman ($\rho$KR) term can
contribute to these isospin channels in addition
to four diagrams appearing in the $\gamma N \rightarrow \pi^{+} \pi^{-} N$.
The $\rho$KR term [Fig. 1(e)] and the $N^*$ terms
dominate in this case, and the interference among the $\rho$KR term,
$N^*\rho$N term and the $\Delta$KR term is important and gives rise to the
bump in the excitation curve as shown in Figs. \ref{fig:pzn}(a) and (b).
With regards to the  $\gamma N \rightarrow \pi^{0} \pi^{0} N$ reaction,
the magnitude of the cross section is underestimated about a factor of
$\frac{3}{4}$ in our model. However, the underestimate of this channel does
not
affect our conclusion as far as the total cross section is concerned,
since the cross section of
$\gamma N \rightarrow \pi^{0} \pi^{0} N$ reaction is smaller than 10
percent of total two-pion production reaction.

 What we learned from the elementary processes are the followings.
\begin{enumerate}
\item  The cross section of one-pion photoproduction has only a small
       bump for a proton and a shoulder for a neutron in the $N^*$
       resonance energy region. Therefore, we can easily make the $N^*$
       resonance peak from the one-pion production vanish by introducing
       a much smaller width due to the collision broadening than those
       given by Alberico et al.\cite{Phe1} and Kondratyuk et al.\cite{Phe2}.
\item  For the two-pion photoproduction the $N^*$ contribution alone
       is not large. In order to give rise to the bump in the cross
       section, the interference between the $N^*$ term and other terms
       is very important. So, we expect that the delicate balance of the
       interference is broken in the nuclear medium by the collision
       broadenings of $\Delta$ and $N^*$, the pion distortion in the
       $\pi \Delta$ channel and the Fermi motion, and therefore,the
       bump is strongly suppressed due to cooperative effects of the
       broadenings and interference.

\end{enumerate}
\begin{figure}[t]
\begin{center}
\includegraphics[height=10cm,width=12cm]{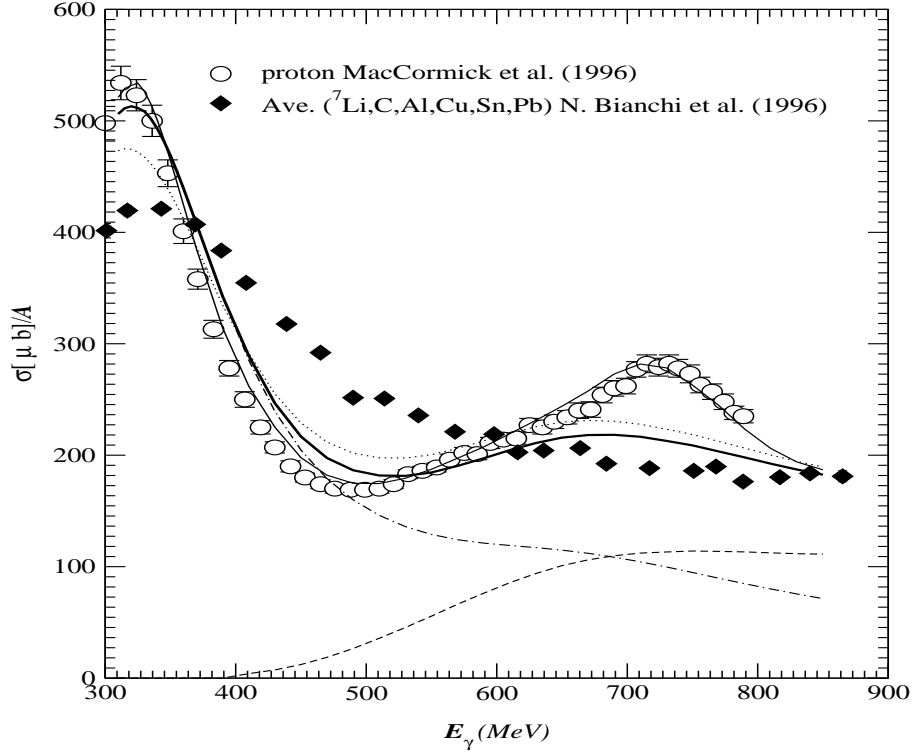}
\end{center}
\vspace{-1.0cm}
\caption{\small{
 The Fermi motion effects for total nuclear photoabsorption cross section on
nuclei.
The dotted line is the nuclear cross section averaged over the initial
nucleon momentum. The thick solid line is the nuclear cross section
including the
Pauli blocking effect for the final emitted nucleon and intermediate
$\Delta$
state in addition
to the average over the initial nucleon momentum. The dash-dotted
and dashed lines are
two components of the thick solid line,i.e.,the one-pion and two-pion
production,
respectively. The thin solid line corresponds to the cross section
on a free proton calculated by using multipole amplitudes.
Experimental data for nuclei are taken from
Ref. \protect\cite{Exp3}. The open circles represent data of the total
photoabsorption cross
section on a proton from Ref.\protect\cite{Exp5}.
}}
\label{fig:fermi}
\end{figure}

 Now we discuss the total cross section of nuclear photoabsorption.
For simplicity we adopt the Fermi gas model for a nucleus, and
${k^{av}}_f = \int d\vec{r}\rho (\vec{r})k_f(\rho)$ as the Fermi momentum
in order to take into account the finiteness of nucleus. In our calculation,
the total cross section per nucleon is obtained by taking the average
of contributions from a proton and neutron in the nuclear matter.

As shown in Fig. \ref{fig:fermi}, the Fermi motion produces strong
damping of the cross section around the $N^*$ resonance energy region.
However, the small bump still remains and its effect cannot fill up the
valley
between 380 and 500 MeV.  For comparison, we also show the experimental
cross
section on a free proton with the theoretical curve in Fig. \ref{fig:fermi}.
Furthermore, to see the details of the Fermi motion effects,
individual contributions for the one-pion and two-pion productions are
presented in Fig. \ref{fig:fermi}. The size of
the Pauli blocking effect for the intermediate and final states is found
to be small but non-negligible as is seen from the difference
between the dash-two-dotted and thick solid lines.

To explain the data, thus, one inevitably needs the other damping
mechanisms.
As additional medium corrections, we take into account the spreading
potentials
\cite{Theo4} for the $N^*$ and $\Delta$ resonances
, and the pion distortion appearing in the formula derived in previous section.
\begin{figure}[t]
\begin{center}
\includegraphics[height=10cm,width=12cm]{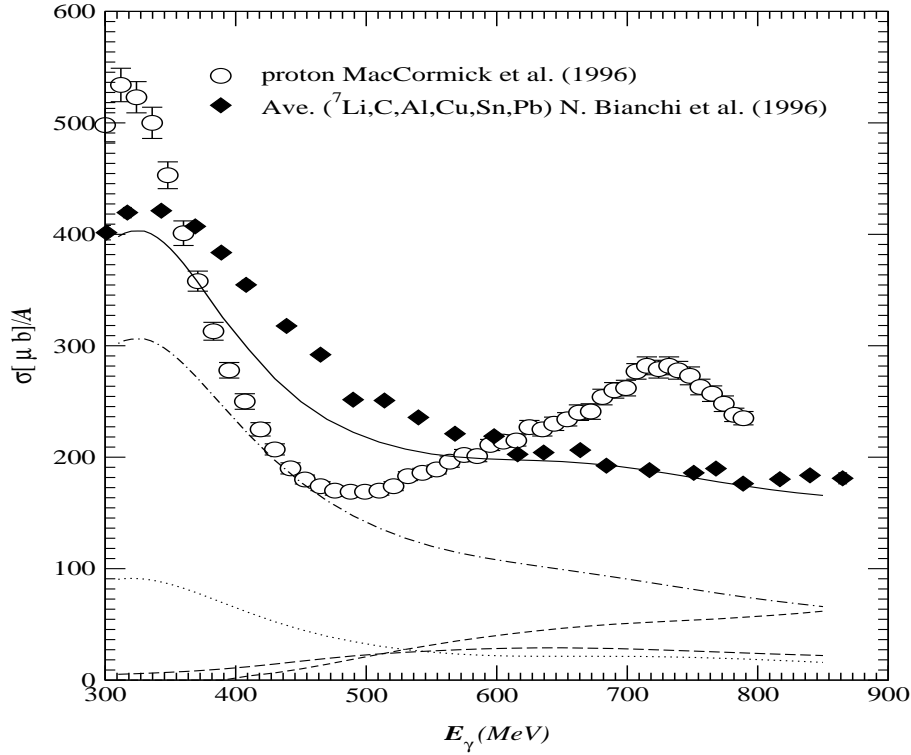}
\end{center}
\vspace{-0.5cm}
\caption{\small{
The total nuclear photoabsorption cross section on nuclei.
The thick solid line is the full calculation. The dash-dotteded and dashed
lines are
the contributions of the one-pion and two-pion production, respectively.
The dotted line is the contribution for the processes of Figs. 1(f)
and
1(g). The long dashed line is the contribution for the process of
Fig. 1(h).
Experimental data for nuclei are taken from Ref.\protect\cite{Exp3}.
The open circles represent data of the total photoabsorption cross
section on a proton from Ref.\protect\cite{Exp5}.}}
\label{fig:nuclear-tot}
\end{figure}
The mass shift and collision width of $\Delta$ have been
already known in the studies of pion-nucleus scattering using the $\Delta$
-hole model \cite{hirata,moniz,lenz,oset} where they can
be identified as the
spreading potential. The spreading potential found in these studies is
almost energy independent. We take $\delta M_{\Delta}=6$ MeV and
$\Gamma_{\Delta sp}=36$ MeV. As the pion self-energy, we adopt the pion
optical potential used by Arima et al. \cite{arima}. As for the mass
shift and collision width of $N^*$, there are no establisfed values
at present. For simplicity, we assume that $\delta M_{N^*}$ and
$\Gamma_{N^*sp}$ are energy independent like $\delta M_{\Delta}$ and
$\Gamma_{\Delta sp}$. Then we vary the values so that the
total nuclear photoabsorption cross sections from 600 to 800 MeV are
reproduced. We found $\delta M_{N^*}=12$ MeV and $\Gamma_{N^*sp}=48$
MeV.

The total photoabsorption cross sections per nucleon (solid line) calculated
with the above parameters are shown in Fig. 6.
It is found that the simultaneus inclusion of the spreading potentials
for the $N^*$ and $\Delta$ resonances
, and the pion distortion gives rise to the complete suppression of the bump
around the $N^*$ resonance energy region. To see the detailed contents of
our calculations,furthermore, we show
each contribution of the absorption processes : the one-pion production
(dash-dotted line), the two-pion production (dashed line), the many-body
absorption through the $\Delta$-nucleus state and $N^*$-nucleus state
(dotted line) corresponding to
Figs. \ref{fig:fyman-diag}(f) and (g), and
the many-body absorption through the $\pi\Delta$-nucleus state (long dashed
line) corresponding to Fig. \ref{fig:fyman-diag}(h). The correction terms
$\Delta_{1}$ and $\Delta_{2}$ in Eqs.(2.36) and (2.37) are already included
in the calculations of the dashed and long dashed lines, respectively.
The size of the correction terms is small but non-negligible. For instance,
there is about 20 percent effect at 750 MeV for the two-pion production.
In the one-pion
photoproduction, the bump near the mass of the $N^*$ disappears by the
spreading potential for $N^*$. The cross section of the two-pion
photoproduction is about 3 times smaller than that of the elementary
process by the cooperative effects between the following medium
corrections : the spreading potentials for $\Delta$ and $N^*$, the pion
distortion, and the change of the interference among the related reaction
processes. The cross sections of the other many-body processes are almost
flat in the energy range above 600 MeV and small. As a consequence of
these effects, the excitation peak around the position of the $N^*$
resonance in the total nuclear photoabsorption cross section disappears
differently from the hydrogen.

Our model, however, underestimates cross
sections in the valley region between 380 and 500
MeV by about 15 percent ($\sim$ 45 $\mu$b). There must be some important
processes which give enhancement for the nuclear photoabsorption in the
\begin{figure}[t]
\begin{center}
\includegraphics[height=10cm,width=12cm]{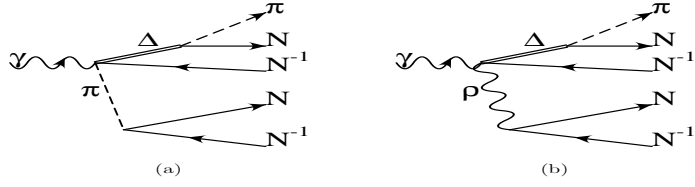}
\end{center}
\vspace{-7cm}
\caption{\small{
The diagrams of two body contribution in the one-pion production.}}
\label{fig:two-body}
\end{figure}
valley region but do not appear in the photoabsorption off a nucleon.
The candidates for such processes are
shown in Figs. \ref{fig:two-body}(a) and (b) where the intermediate pion
and $\rho$ meson are far off-shell. Two nucleons explicitly
contribute to these processes. These contributions are suitable to
explain the mass number dependence $A^{1.7}$ of total cross section in
the valley region. In Refs.\cite{Phe3,carrasco} the contribution from
Fig. \ref{fig:two-body}(a) is taken into account to increase the cross
section of one-pion production.

 The cross sections are also underestimated slightly at the $\Delta$
resonance energy around 320 MeV. This missing strength may be due to
the coherent $\pi^0$ production mechanism in addition to the above
two-nucleon mechanism. The coherent $\pi^0$ production is not included
in our calculation using the Fermi gas model. In this energy region,
our model has to be extended so as to treat the finiteness of the
nucleus more reliably,as was done by Koch et al.\cite{KMO}.

\section{Conclusions}
\label{sec:Conclusions}

 The formula derived by using the projection method for the
photonuclear total absorption cross section has been presented.
Our method is very effective for the case that the
interference effect in the photoabsorption off a nucleon such as
two-pion productions is strong.

 The disappearance of the peak around the position of the $D_{13}$
resonance in the nuclear photoabsorption can be explained by taking
into account the cooperative effect of the interference in the
two-pion photoproduction processes,the collision broadening of
$\Delta$ and $N^*$,and the pion distortion in the nuclear medium.
The change of the interference by the medium plays an important role.
The mass shift and collision width of $N^*$ are found to be
$\delta M_{N^*}=12$ MeV and $\Gamma_{N^*sp}=48$ MeV,respectively.
The collision width obtained is about 6 times smaller than those
in Refs.\cite{Phe1,Phe2}. The total absorption cross sections in
our theoretical calculation around 320 MeV are about
5 percent smaller than average experimental cross sections for several
nuclei. Furthermore,theoretical total absorption cross sections in the
valley
region between 380 and 500 MeV are about 15 percent smaller than
the average experimental cross sections. In this energy region it may be
necessary to take into account such reaction processes as
Figs. \ref{fig:two-body}(a) and (b)
involving two nucleons explicitly.

\newpage

\end{document}